\documentclass[%
 reprint,
 amsmath,amssymb,
 aps,
]{revtex4-1}

\usepackage{graphicx}
\usepackage{dcolumn}
\usepackage{bm}
\usepackage{subfigure}
\usepackage{xcolor}

\def\bc{\begin{center}}
\def\ec{\end{center}}
\def\be{\begin{eqnarray}}
\def\ee{\end{eqnarray}}

\begin{document}

\preprint{APS/123-QED}

\title{Constraints on Bose-Einstein-condensed Axion Dark Matter \\from The H{\small I} Nearby Galaxy Survey data}

\author{Ming-Hua Li}
  \email{limh@ihep.ac.cn;}
\author{Zhi-Bing Li}%
 \email{stslzb@mail.sysu.edu.cn}
\affiliation{%
 School of Physics and Engineering, Sun Yat-Sen University, Guangzhou 510275, China
}%




\date{\today}

\begin{abstract}
One of the leading candidates for dark matter is axion or axion-like particle in a form of Bose-Einstein condensate (BEC). 
In this paper, we present an analysis of 17 high-resolution galactic rotation curves from ``The H{\footnotesize I} Nearby Galaxy Survey (THINGS)'' data [F. Walter et al., Astron. J. 136, 2563 (2008)] in the context of the axionic Bose-Einstein condensed dark matter model.
Assuming a repulsive two-body interaction, we solve the non-relativistic Gross-Pitaevskii equation for $N$ gravitationally trapped bosons in the Thomas-Fermi approximation. We obtain the maximum possible radius $R$ and the mass profile $M(r)$ of a dilute axionic Bose-Einstein condensed gas cloud.  
A standard least-$\chi^2$ method is employed to find the best-fit values of the total mass $M$ of the axion BEC and its radius $R$. The local mass density of BEC axion dark-matter is $\rho_{a}\simeq 0.02~{\rm GeV/cm}^3$, which agrees with that presented by Beck [C. Beck, Phys. Rev. Lett. 111, 231801 (2013)].  The axion mass $m_a$ we obtain depends not only on the best-fit value of $R$ but also on the $s$-wave scattering length $a$ ($m_a \propto a^{1/3}$). The transition temperature $T_a$ of axion BEC on galactic scales is also estimated.  Comparing the calculated $T_a$ with the ambient temperature of galaxies and galaxy clusters implies that $a\sim 10^{-3}$ fm. The corresponding axion mass is $m_a\simeq 0.58$ meV. We compare our results with others. 
\begin{description}
\item[PACS numbers]
67.85.Jk; 95.35.+d; 98.62.Ck; 98.62.Dm
\end{description}
\end{abstract}

\pacs{Valid PACS appear here}
\maketitle


\section{\label{sec:level1}Introduction}

Astronomical observations in the past few decades have all indicated the existence of a cold, collisionless, non-baryonic substance in our universe: the Oort discrepancy \cite{Oort:1932}, the asymptotic behavior of galactic rotation curves \cite{Binney1987, Persic1996, Walter2008}, the Bullet Cluster and the cluster gas masses \cite{Clowe2007}, the gravitational lensing observations \cite{Takahashi2007}, the structure formation \cite{Caldera-Cabral2009}, etc. The substance has been dubbed ``dark matter'' since the essence of it still remains obscure. Besides weakly interacting massive particles (WIMPs) \cite{Bertone2010} and sterile neutrinos \cite{Dodelson1994, Seljak2006}, axion is a leading candidate for dark matter \cite{Mielke2009, Hwang2009}. 

Axion is a pseudo-Nambu-Goldstone boson \cite{Weinberg1978, Wilczek1978} resulting from spontaneously breaking the Peccei-Quinn symmetry \cite{Peccei1977}. It has first been invented to suppress the strong charge-parity (CP) violation in Quantum Chromodynamics (QCD).  Axion is different from other dark matter candidates like WIMPs and sterile neutrinos because it can form a Bose-Einstein condensate (BEC). It has been postulated that low mass axions could form a BEC of astronomical extent, which could plausibly explain the missing mass problem of our universe \cite{Sin1994, Silverman2002}. Lots of work have done in the past few years about axionic dark matter and its astronomical implications. To name a few: axionic BECs are possible to form lump-like dark matter structures \cite{Avelar2008} similar to those of boson stars \cite{Schunck2003};
\citet{Sikivie2009} proposed that BEC dark-matter axions, which continually falling into a galaxy, give rise to ripple-like fine structures in galactic rotation curves called ``caustic rings'' \cite{Sin1994, Duffy2008, Banik2013}; Harko claimed that BEC dark matter solves the core/cusp problem of rotation curves for a sample of eight dwarf galaxies \cite{Harko2011}; 
Lee and Lim proposed that BEC/SF dark matter theory can explain the minimum length scale and the minimum mass scale (also their dependence on the brightness) of dwarf galaxies \cite{Lim2010} ; 
Erken {\it et al.} suggested in their paper that BEC dark matter provides a possible mechanism for photon cooling after big bang nucleosynthesis but before recombination, which resolves the Lithium problem \cite{Erken2012}, etc.

Observational detections of axionic BEC dark matter are underway. A promising way to search for axion or axion-like particle is its conversion into two photons in an external magnetic field \cite{Ehret2010}. Relevant observations include the Axion Dark Matter Experiment (ADME) \cite{Duffy2006} and the CERN Axion Solar Telescope (CAST) \cite{Arik2013}. An upper bound of the axion-photon coupling strength $g_{a\gamma\gamma} \leq 3.3 \times 10^{-10} ~{\rm GeV}^{-1}$ and the possible axion mass range for future searching have also been obtained. 
In theoretical studies, the axion mass is actually a free parameter with its possible range from a few $\mu$eV to a few eV.
To answer the questions such as whether we could discover it in our next-generation detector, or whether it could always stay thermalized throughout the history of our universe, one has to know the mass of an axion particle.

As a supplement to these pioneering works, in this paper, we try to estimate the local mass density of the axionic dark matter from astronomical observations, especially on the galactic scale. The flatness of rotation curves of spiral galaxies is the first and one of the most convincing evidence for the existence of dark matter. It is therefore a mandatory test for a whatever new dark matter candidate. From the galactic rotation curves and velocity dispersions, one can also place constraints on the local mass density and other properties of dark matter \cite{Bosch2001,Blok:2008}. We present an analysis of 17 high-resolution galactic rotation curves from ``The H{\footnotesize I} Nearby Galaxy Survey (THINGS)'' \cite{Walter2008} in the context of an axionic BEC dark matter model. We consider a dilute, repulsive axionic gas cloud which is self-gravitationally trapped. The generalized non-relativistic Gross-Pitaevskii (GP) equation is solved to obtain the maximum possible radius and the mass profile of the axionic BEC cloud. The radius $R$ and the total mass $M$ of the dark-matter axions in a galaxy are taken as free parameters in the numerical analysis. Given the best-fit values of $R$ and $M$, the single-particle mass $m_a$, the mass density $\rho_a$, and the transition temperature $T_a$ of the dark-matter axions are deduced.

The rest of the paper is organized as follows. In Section II, we solve the generalized GP equation in the Thomas-Fermi approximation to obtain the optimal form of the wave function of the axionic BEC state. In Section III, we calculate the mass profile of the dark-matter axions and the tangential velocity for each of the sample galaxies. We compare the theoretical predictions with the THINGS data using a standard least-$\chi^2$ method. The best-fit values and the $1$-$\sigma$ (68.3\% C.L.) error of the parameters $M$ and $R$ are presented. In Secion IV, we deduce the corresponding mass density and the critical temperature of the axionic BEC dark matter. We compare our results with those of others. Discussion and conclusions are given in Section V. A pedagogical derivation of the generalized GP equation and its solution is given in Appendix.

\section{The generalized Gross-Pitaevskii equation and the Thomas-Fermi approximation}\label{sec:theory}
A BEC is a state of dilute bosonic gas which all the bosons ocuppy the lowest quantum state. To attain such a state, the system has to be cooled down to textcolor{blue}{a certain temperature}. In this state, the bosons behave like a large ``atom'', displaying macroscopic quantum effects. This state was first prepared in the laboratory from Rb vapour in 1995 \cite{Anderson1995, Davis1995} about seventy years after it was theoretically predicted by \cite{Einstein1925}.

To study the axionic BEC of astronomical scale, one has to solve the Gross-Pitaevskii equation, which describes the properties of the non-uniform bosonic gas while the wavelength is larger than the distance between two atoms. We first derive the generalized Gross-Pitaevskii (GP) equation by variational methods, considering a repulsive interactions between axions. Then we solve the equation under the Thomas-Fermi approximation for dilute gas, in which the kinetic energy term is neglected. In next section, we use the result to calculate the tangential velocity contributed by the axionic BEC dark matter to the total rotation velocity of a galaxy.

The wave function for a system of $N$-identical bosons is symmetric under the interchange of any two particles. So the wave function of an $N$-boson system is given as 
\be
\Psi({\bf r_1}, {\bf r_2}, ..., {\bf r}_N)=\prod_{i=1}^N \phi({\bf r}_i),
\label{wavefuncN}
\ee
where $\int  d{\bf r} |\phi({\bf r})|^2=1$. $\phi({\bf r})$ is the normalized wave function of a single boson.
The wave function of the condensed state is defined as
\be
\psi({\bf r})=N^{1/2} \phi({\bf r}).
\label{psi}
\ee
The number density of particles is given by \cite{Pethick2008}
\be
n({\bf r})=|\psi({\bf r})|^2.
\ee

For a dilute axionic gas cloud which is self-gravitationally trapped, the Hamiltonian of the system can be written as
\be
H=\sum_{i=1}^N \left[\frac{{\bf p}_i^2}{2m}  + V({\bf r}_i) \right]+ b \sum_{i<j} \delta({\bf r}_i-{\bf r}_j),
\label{hamiltonian}
\ee
where
\be
V({\bf r})=-\frac{GMm}{r}\int_0^r |\phi({\bf r'})|^2 d{\bf r'}
\ee
is the gravitational potential energy of the gas cloud and $M=Nm$ is the total mass of the cloud. $m$ is the mass of a single-particle. 
The effective interaction between two particles is described by
$b \delta({\bf r}_i-{\bf r}_j)$, for which a positive $b$ means that the atom-atom interaction is repulsive.
The energy of the state is given by
\begin{widetext}
\be
E=\langle \psi| {\hat H} |\psi \rangle=N \int d{\bf r} \left[-\frac{\hbar^2}{2m} |\nabla \phi({\bf r})|^2  -\frac{GMm}{r}\int_0^r |\phi({\bf r'})|^2 d{\bf r'}|\phi({\bf r})|^2 + \frac{N-1}{2} b |\phi({\bf r})|^4 \right],
\label{E}
\ee
\end{widetext}
To find the optimal form of the wave function of the condensed state $\psi({\bf r})$, one has to minimize the energy (\ref{E}) with respect to the variation of $\phi({\bf r})$ or its conjugate $\phi^{\ast}({\bf r})$. 
We use the Lagrange multiplier method to take into account of the constraint that the total particle number $N$ is constant, i.e. $\delta (E-\mu \int d{\bf r}|\psi({\bf r})|^2  )=0$. The Lagrange multiplier $\mu$ can be interpreted as the chemical potential of the $N$-boson system. 
After some textbook calculations,  one obtains the GP equation for a dilute axionic gas cloud which is self-gravitationally trapped, 
\begin{widetext}
\be
-\frac{\hbar ^2}{2m}\nabla^2 \phi(r) + \left[-\frac{4\pi GM}{r}\int_0^r |\phi(r')|^2 r'^2 dr' -4\pi  \int_r^\infty \frac{GM}{r'}|\phi(r')|^2 r'^2dr'+(N-1)b|\phi(r)|^2\right]\phi(r)=\mu \phi(r).
\label{GP1}
\ee
\end{widetext}

Before solving equation (\ref{GP1}), we apply the Thomas-Fermi approximation for a dilute gas and drop the kinetic term. By introducing the radial wave function $u(r)\equiv r\phi(r)/Y_{00}$ ($Y_{00}=1/\sqrt{4\pi}$ is the spherical harmonic function), equation (\ref{GP1}) can be rewritten as
\begin{widetext}
\be
\left[-\frac{GM}{r}\int_0^r |u(r')|^2dr' -\int_r^\infty \frac{GM}{r'}|u(r')|^2 dr'+ \frac{(N-1)}{4\pi}b\left|\frac{u(r)}{r}\right|^2\right]u(r)=\mu u(r),
\label{GP2}
\ee
\end{widetext}
Given that $b=4\pi \hbar^2 a/m$ \cite{Pethick2008}, where $a$ is the $s$-wave scattering length, the optimal form of the wave function of the condensed state can be obtained by solving equation (\ref{GP2}), to wit 
\be
|\phi(r)|^2=\frac{1}{4R^2}\frac{{\rm sin}\left(\frac{\pi r}{R}\right)}{r}, ~~~r\in [0, R],
\label{phi2}
\ee
where
\be
R=\pi\sqrt{\frac{\hbar^2 a}{Gm^3}} ,~~~~{\rm for}~N\gg 1
\label{radius}
\ee
is the cut-off radius, since in physics one has to ensure $|\phi(r)|^2\geqslant 0$. The $s$-wave scattering length $a$ is defined as the zero-energy limit of the scattering amplitude, the value of which is given by some terrestrial laboratory experiments \cite{Dalfovo1999}.
This result is the same as that obtained by \cite{Bohmer2007}. But we emphasize that as seen in equation (\ref{radius}), for a certain kind of bosonic particle,  such as axions, which has a certain single-particle mass $m$ and a definite $s$-wave scattering length $a$, the cut-off radius $R$ is determined. That places an upper bound on the size of the BEC cloud formed by axions or other axion-like particles. In next section, we will use the galactic rotation curves data to determine the value of $R$.

\section{Mass Profile and Galactic Rotation Curves}\label{sec:rotation curves}
As mentioned in the introduction, the flatness of spiral galaxy rotation curves is a compelling test for whatever a dark matter model. So in this section, we deduce the mass profile of the axionic BEC and then use it to calculate the rotation velocity for a spiral galaxy. Numerical study is carried out to find best-fit values of the radius $R$ and the total mass $M$ of the BEC axion core.

\subsection{The BEC Mass Profile}
The mass density of a BEC is  
\be
\rho({\bf r})=mn({\bf r})=m|\psi({\bf r})|^2.
\label{rho}
\ee
Given equation (\ref{psi}) and (\ref{phi2}), we obtain the mass density profile of the axionic BEC dark matter,
\be
\rho_{{\rm BEC}}(r)=\frac{M}{4R^2}\frac{{\rm sin}\left(\frac{\pi r}{R}\right)}{r}, ~~~r\in [0, R],
\label{massdensity}
\ee
where $M$ is the total mass of the axion BEC cloud.
The mass profile of the axionic BEC dark matter is given as
\be
M_{{\rm BEC}}(r)&=&4\pi \int_0^r r'^2\rho_{{\rm BEC}}(r')dr' \nonumber \\ 
&=&M\left[\frac{1}{\pi}{\rm sin}\left(\frac{\pi r}{R}\right)-\frac{ r}{R}{\rm cos}\left(\frac{\pi r}{R}\right)\right] 
\label{mass2}
\ee
for $0\leq r \leq R$.

\subsection{Galactic Rotation Curves}
The rotation velocity of a galaxy is given as \cite{Swaters2010}
\be
v_{\rm th}(r)=\sqrt{v_{\ast}^2(r)+v_{\rm gas}^2(r)+v_{\rm BEC}^2(r)},
\label{velocity}
\ee
where $v_{\ast}(r)$, $v_{\rm gas}(r)$ and $v_{\rm BEC}(r)$ represents the contribution of the stellar disk, the interstellar medium (ISM) gas, and the axion BEC respectively. $v_{\rm BEC}$ is calculated from equation (\ref{mass2}) by 
\be
v_{\rm BEC}(r)&=&\sqrt{\frac{GM_{\rm BEC}(r)}{r}} \nonumber \\
&=&\sqrt{\frac{GM}{r}\left[\frac{1}{\pi}{\rm sin}\left(\frac{\pi r}{R}\right)-\frac{ r}{R}{\rm cos}\left(\frac{\pi r}{R}\right)\right]}.
\label{vdark matter}
\ee

The velocity profile of stellar disk $v_{\ast}(r)$ is derived from the photometric data of a galaxy. According to \cite{Vaucouleurs1959}, the surface brightness of the stellar disk can be well fitted by an exponential law 
\begin{equation}\label{eq:brightness}
  I(r)=I(0)e^{-r/h},
\end{equation}
where $h$ is the scale length and $I(0)$ is the surface brightness at the center of galaxy. $r=\sqrt{x^2+y^2}$, where the $x$- and $y$-axis lie in the galactic plane.  
Assuming a mass-to-light ratio constant with radius of a galaxy and an infinitely-thin disk, the mass profile of the stellar disk is given as
\begin{equation}
  \Sigma(r)=\Sigma(0)e^{-r/h},
\end{equation}
where $\Sigma(0)=\Upsilon_{*}I(0)$, $\Upsilon_{*}$ is the stellar mass-to-light ratio of a galaxy.
The total mass of the stellar disk is given by
\begin{equation}
\label{eq:disk-mass}
 M_{\ast}=\int_0^{\infty} 2\pi r \Sigma(r)dr=2\pi\Sigma(0)h^2.
\end{equation}
 By similarly integrating equation (\ref{eq:brightness}), one gets the total luminosity $L$ of a galaxy,
 \begin{equation}
 L=2\pi I(0)h^2.
\label{L2}
\end{equation}

The mass-to-light ratio of a galaxy can therefore be defined as
\begin{equation}\label{eq:mass-to-light}
  \Upsilon_{\ast}=\frac{M_{\ast}}{L}=\frac{\Sigma(0)}{I(0)}.
\end{equation}
$ \Upsilon_{\ast}$ is usually taken as a free parameter to be determined by comparing theoretical predictions with observations. In next section, we take $\Sigma(0)$ instead of $ \Upsilon_{\ast}$ to be a free parameter. From equation (\ref{eq:mass-to-light}), one can see that these two are equivalent, provided that $I(0)$ is known. In fact, the total luminosity $L$ of a galaxy is related to its absolute magnitude 
$\mathcal{M}$ via
\begin{equation}\label{eq:M-L-relation}
  \mathcal{M}-\mathcal{M}_{\odot}=-2.5{\rm log}_{10}\frac{L}{L_{\odot}},
\end{equation}
where $\mathcal{M}_{\odot}$ and $L_{\odot}$ respectively represents the absolute magnitude and total luminosity of the sun. 
The central surface brightness $I(0)$ can be obtained by substituting equation (\ref{L2}) into (\ref{eq:M-L-relation}) provided that $\mathcal{M}$ and $h$ of a galaxy are given by the observations. Thus, we use $\Sigma(0)$ instead of $\Upsilon_{\ast}$ as a free parameter to do the numerical analysis.

The rotation velocity contributed by the exponential disk is given by \cite{Freeman1970}
\begin{equation}\label{eq:disk-velocity}
  v_{\ast}(r)=\sqrt{\frac{M_{\ast}}{r}\gamma(r)},
\end{equation}
where
\begin{equation}
  \gamma(r)\equiv\frac{r^3}{2h^3}\left[I_0\left(\frac{r}{2h}\right)K_0\left(\frac{r}{2h}\right)- I_1\left(\frac{r}{2h}\right)K_1\left(\frac{r}{2h}\right)\right].
\label{gamma}
\end{equation}
Here, $I_{n}$ and $K_{n}$ ($n=0,~1$) are respectively the $n$th-order modified Bessel functions of the first and second kind. $M_{\ast}$ in equation (\ref{eq:disk-velocity}) is given by equation (\ref{eq:disk-mass}).

The contribution of ISM is given by
\begin{equation}\label{eq:newton-velocity}
  v_{\rm gas}(r)=\sqrt{\frac{4}{3}v_{\rm HI}^2(r)},
\end{equation}
where $v_{\rm HI}(r)$ is read directly from the photometric data. The factor 4/3 in equation (\ref{eq:newton-velocity}) comes from the contribution of both helium (He) and neutral hydrogen (HI). Here we assume that the mass ratio of He and HI is $M_{\rm He}/M_{\rm HI}=1/3$. Any other gases are negligible compared to HI and He.
Therefore, equation (\ref{velocity}) can be rewritten as 
\begin{equation}\label{eq:fit-formula}
  v(r)=\sqrt{\frac{4}{3}v_{\rm HI}^2(r)+v_{*}^2(r)+v_{\rm BEC}^2(r)}.
\end{equation}
where $v_{\rm BEC}(r)$ and $v_{*}(r)$ are given by equation (\ref{vdark matter}) and (\ref{eq:disk-velocity}).

\subsection{Numerical Analysis}
In this section, we make fits by equation (\ref{eq:fit-formula}) to the observed spiral galaxy rotation curves data with a least-squares algorithm. We employ ``The H{\footnotesize I} Nearby Galaxy Survey (THINGS)'' dataset \cite{Walter2008} to do the numerical analysis. It contains 17 galaxies with the highest quality rotation curves data currently available. 
The inclinations, distances and absolute magnitudes of these galaxies have all been measured with high precision. The sample galaxies and their properties are listed in Table \ref{tab:samples}. 

\begin{figure*}
 \includegraphics[width=15.5cm,height=21cm]{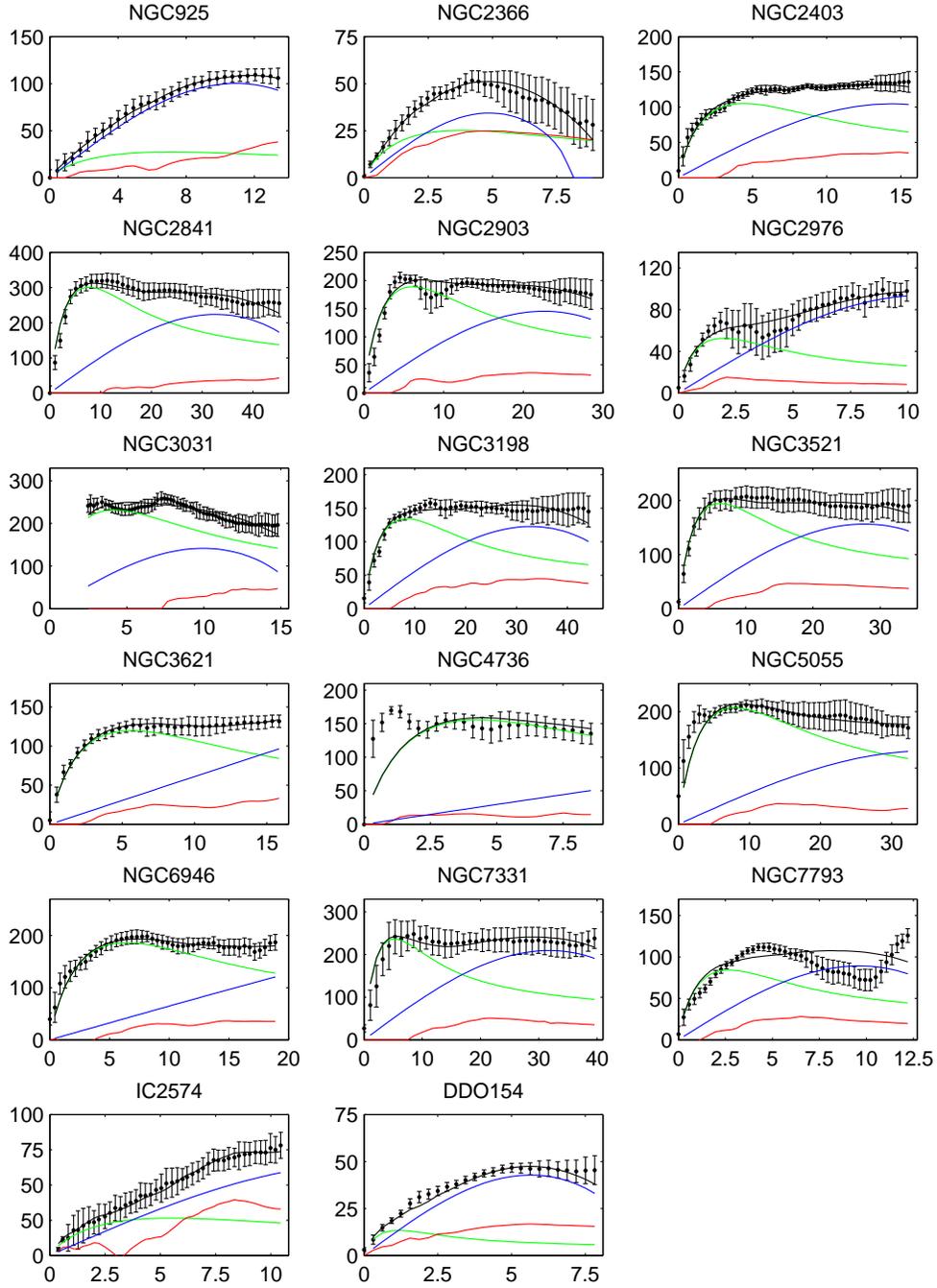}
  \caption{\small{Three-parameter fit (black curves) to the rotation curves of sample galaxies. The $x$-axis is the radial distance in kpc, and the $y$-axis is the rotation velocity in km~s$^{-1}$. The black solid curves indicate the theoretically predicted total rotation velocity calculated from equation (\ref{velocity}). The blue solid curves are calculated from the axion BEC dark matter by equation (\ref{vdark matter}). The red solid curves are the contribution of gas (HI and He). The green curves are the contribution of the stellar disk in Newtonian dynamics.}}
  \label{fig:figure1}
\end{figure*}

The free parameters in our fits are: central surface mass density $\Sigma(0)$, total axionic BEC mass $M$, and cut-off radius  $R$ of the axionic BEC cloud. We use a standard least-$\chi^2$ method, first defining the Chi-square as
\begin{equation}\label{eq:chi-aquare}
  \chi^2=\sum_{i=1}^n \frac{[v_i^{\rm obs}-v^{\rm th}(r_i)]^2}{\sigma_i^2},
\end{equation}
where $v^{\rm th}(r_i)$ is the theoretically predicted tangential velocity at radius $r_i$ calculated from equation (\ref{eq:fit-formula}). 
To deduce ISM gas contribution $v_{\rm HI}(r_i)$ in equation (\ref{eq:fit-formula}), we employ the method proposed by Begeman \cite{Begeman1989} that is used commonly in other literatures \cite{Sanders1998,Verheijen2001,Sanders2007}. We first use the task \textsc{radial} \cite{radial} of GIPSY \cite{gipsy} to derive the radial HI surface density distribution from the 2-dimensional integrated HI maps (robust weighted moment-0) from the THINGS data cube. Then we run the task \textsc{radprof} \cite{radprof} to deduce the contribution to the rotation curve from the obtained HI surface density profile, i.e. $v_{\rm HI}(r_i)$.
The observed  rotation velocity $v_i^{\rm obs}$ is deduced from the THINGS data cube (robust weighted moment-1) using the task \textsc{rotcur} of GIPSY, with PA, INCL, $V_{\rm sys}$ and galaxy center (RA and DEC) fixed at the values given in Table \ref{tab:samples}. $\sigma_i$ is the uncertainty  of $v_i^{\rm obs}$. We minimize equation (\ref{eq:chi-aquare}) to find the best-fit values of these parameters. The results are presented in Figure \ref{fig:figure1}, and the best-fit parameter values with $1$-$\sigma$ (68.3\% C.L.) error are listed in Table \ref{tab:parameters1}. 

Notwithstanding the high quality circular velocity measurements, not all of the galaxies in the THINGS sample yield a good result in the fits. For NGC3621 and NGC4736, a best-fit value of the parameters is obtained but with uncertain error. For NGC4736, this is possibly because of the poor quality of the data --- the observed rotation velocity extends to only about 7 kpc and wiggles near the galactic center $r=0$. For NGC3031, the tangential velocities for $r<3$ kpc are poorly determined, giving rise to a large error of $\Sigma_0$ and a large $\chi^2$. For NGC3621, this may be attributed to the fact that it shows a trend of ascending at the far end of the observed rotation curve. The error of $M$ and $R$ for NGC6946 and the error of $\Sigma_0$ of NGC3031 and NGC7331 are not very well determined by the fit either possibly because of the same reason. A more detailed numerical analysis is necessary for these galaxies. But that should be the subject of another paper.

\begin{figure*}
\centering
\subfigure[~\textsf{Distribution of $R$}] { \label{fig:a}
\scalebox{0.55}[0.55]{\includegraphics{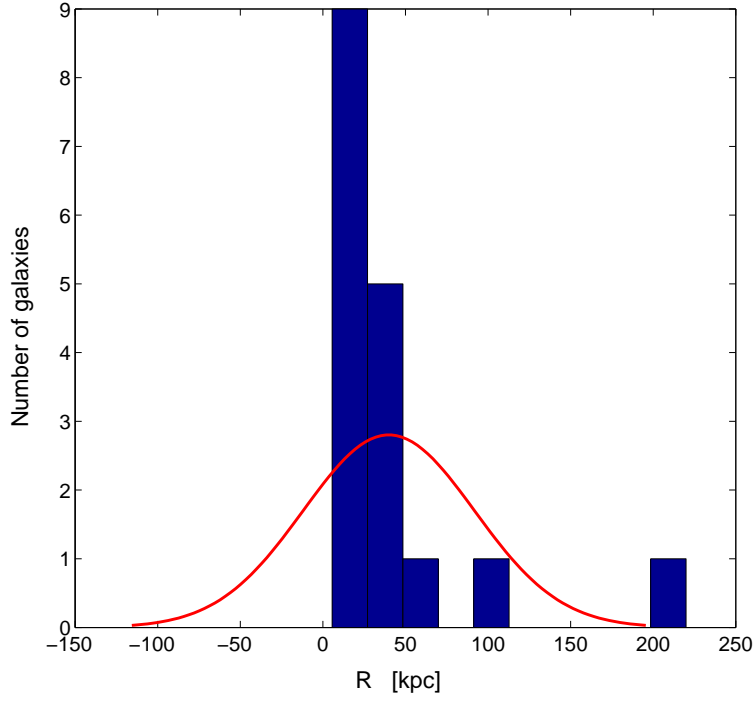}}
}
\subfigure[~\textsf{Distribution of $R$ over $M/L$}] { \label{fig:b}
\scalebox{0.5}[0.5]{\includegraphics{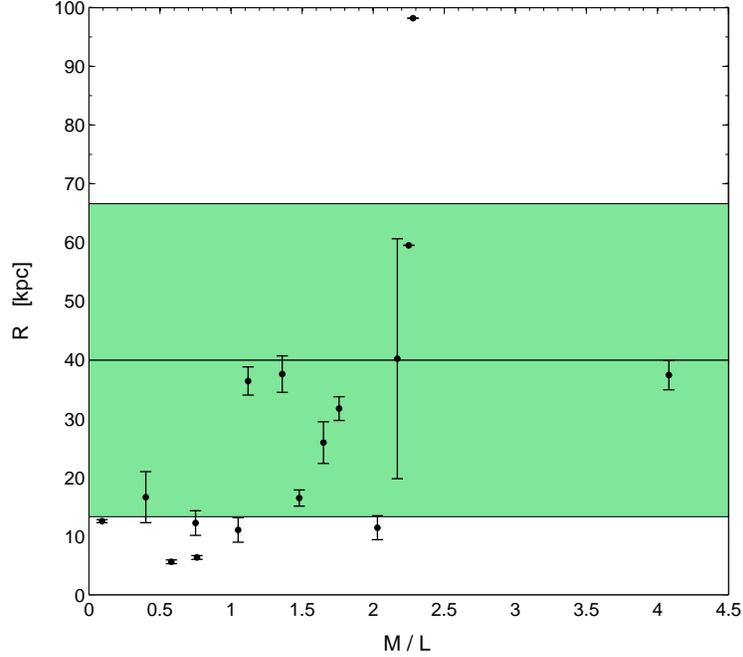}}
}
 \caption{\small{(a) Distribution of the best-fit values of $R$. The best-fit values of $R$ has a range $0\sim 220$ kpc, which is divided into ten bins. The $y$-axis is the number of galaxy in each bin. The red solid curve indicates the best-fit normal distribution of the value of $R$. The expectation value and the standard deviation of $R$ in units of kpcs are respectively $\mu_R=39.9$ and $\sigma_R=51.9$; (b) Distribution of  $R$ over the mass-to-light ratio $M/L$. The error bars indicates $2$-$\sigma$ (95\% C.L.) error of $R$ for each sample galaxy. The data point of NGC3621 is neglected in this plot because its best-fit $R$ is much larger than those of other samples. The errors of $R$ for NGC6946 and NGC4736 have been set to be zero for their uncertainties. The best-fit expectation value of $R$ in units of kpcs is $\mu_R=39.9_{-26.7}^{+26.7}$ (68.3\% C.L.). The green shaded area indicates the $1$-$\sigma$ error range of $\mu_R$.}}
\label{fig:figure2}
\end{figure*}

The distribution of the best-fit values of $R$ is shown in Figure \ref{fig:figure2}(a). It is fitted to a probability density function of normal distribution, of which the expectation value is $\mu_R=39.9_{-26.7}^{+26.7}$ (68.3\% C.L.) in units of kpcs and the standard deviation is $\sigma_R=51.89_{-13.24}^{+27.07}$ (68.3\% C.L.).

The distribution of the best-fit values of $R$ over the mass-to-light ratio $M/L$ is shown in Figure \ref{fig:figure2}(b). The best-fit values of the stellar mass-to-light ratio $M/L$ for the sample galaxies range from $0$ to $2.5$. It can be seen from Figure \ref{fig:figure2}(b) that most of them are distributed near 1 (except for NGC2841). The mean value is $\overline{M/L}=1.48$. This result is in accordance with those given in other references \cite{Swaters2003, Brownstein2006, Blok:2008}.

\section{Mass Density and Critical Temperature}
In this section, we use the best-fit $M$ and $R$ to estimate the mass density and critical temperature of the axionic BEC dark matter for the sample galaxies. We then compare our results with those given by other experiments and observations.

To estimate the axion mass $m_a$, we rewrite equation (\ref{radius}) as 
\be
m_a=\left(\frac{\pi^2 \hbar^2 a}{G R^2}\right)^{1/3}\simeq  6.73 \left(\frac{a}{R^2}\right)^{1/3}\times 10^{-2}{\rm eV}.
\label{singlemass}
\ee
where $a$ and $R$ are in unit of fm and kpc respectively. It should be noted that $m_a$ depends not only on $R$ but also on the $s$-wave scattering length $a$. Given that $R=39.9$ kpc and assuming $a \sim 1$ fm, one has $m_a \simeq 5.8$ meV. A more reasonable choice of the value for $a$ would be $a \sim 10^6$ fm, in the light of the results for different atomic species used in the laboratory experiments on BEC ($a = 2.75$ nm for $^{23}$Na \cite{Tiesinga1996}, $a = 5.77$ nm for $^{87}$Rb \cite{Boesten1997}). This would lead to $m_a \simeq 0.58$ eV. Different values of $a$ would give different axion mass $m_a$. We will discuss this point more detailedly in the last section.

The local axionic BEC dark matter density can be estimated directly from the observationally determined $M$ and $R$ as
\be
\rho_{a}=\frac{M}{4\pi R^3/3} ~~\sim~~0.02~{\rm GeV/cm}^3.
\label{axiondensity}
\ee
This result agrees with that obtained by \citet{Beck2013}, which presented that  $\rho_{a}=0.05~{\rm GeV/cm}^3$. The result in equation (\ref{axiondensity}) and that given by \citet{Beck2013} both satisfy the constraint put by \citet{Hoskins2011}, who found that the local density of non-virialized axionic dark matter should not exceeds $0.2~{\rm GeV/cm}^3$.

According to the BEC theory, the critical temperature is given as \cite{Pitaevskii2003}
\be
T_c=\frac{2\pi \hbar^2}{m k_{{\rm B}}}\left(\frac{n}{\zeta(3/2)}\right)^{2/3},
\label{temperature1}
\ee
where $k_B$ is the Boltzmann constant and $\zeta(3/2)$ is the Riemann zeta function, $\zeta(3/2)\simeq 2.61$. $n\equiv  N/(4\pi R^3/3)$ is local number density of axionic BEC dark matter particles, which is given by (using equation (\ref{singlemass}))
\be
n=\frac{M/m_a}{4\pi R^3/3}=\frac{3M}{4\pi}\left(\frac{G}{\pi^2 \hbar^2 R^7 a}\right)^{1/3}.
\label{number density}
\ee 
Equation (\ref{temperature1}) can be rewritten to get the transition temperature of the axionic BEC dark matter, i.e.
\be
T_a&=&\frac{2}{k_{{\rm B}}}\left(\frac{3}{4\zeta(3/2)}\right)^{2/3}(\pi^{17} \hbar^8)^{1/9}\left(\frac{G^5 M^6}{a^5 R^8}\right)^{1/9}\nonumber \\
&\simeq& \left[\frac{M^6}{a^5 R^8}\right]^{1/9} \times 10^{-5} {\rm eV},
\label{temperature2}
\ee
where $a$, $M$ and $R$ are in unit of fm, ${\rm M}_{\odot}$ and kpc respectively. In equation (\ref{temperature2}), the critical temperature of the BEC axion is expressed as a function of $R$, $M$, and $a$.
For galaxies, $M\sim 10^{11}$ ${\rm M}_{\odot}$, $R \sim 10$ kpc. 
Provided that $a \sim 10^6$ fm, one obtains $T_a \sim 10^{-2}$ eV.
It means that near the galaxies, the temperature of the axionic BEC dark matter halo (if it really exists) would be much lower than that of the interstellar medium (ISM) gas of the galaxies (which takes a value of $10^2 \sim 10^3$ eV). This is also the case for the galaxy clusters. For a medium-sized cluster of galaxies, $M\sim 10^{15}$ ${\rm M}_{\odot}$, $R \sim 10^3$ kpc. Assuming $a \sim 10^6$ fm, one has $T_a \sim 10^{-1}$ eV. It is also lower than the temperature of the intracluster medium (ICM) gas of a cluster (for the Bullet Cluster 1E0657-558, the temperature of ICM gas of the main cluster is about $10^4$ eV \cite{Clowe2007}).

A possible reason for these temperature differences is that $a\sim 10^6$ fm (which is obtained from atomic species in the laboratory experiments) is not a proper estimate of $a$ for axions or other possible axion-like dark matter candidates. In fact, if $a$ takes a value of $a\sim 10^{-3}$ fm, from equation (\ref{temperature2}) one gets: $T_a\simeq 10^3$ eV for galaxies (with $M\sim 10^{11}$ ${\rm M}_{\odot}$ and $R \sim 10$ kpc); $T_a\simeq 10^4$ eV for a medium-sized cluster of galaxies (with $M\sim 10^{15}$ ${\rm M}_{\odot}$ and $R \sim 10^3$ kpc). These results are in accordance with the observations of the ISM/ICM gas of galaxies/galaxy clusters, which were mentioned in the last paragraph. Moreover, for $a\sim 10^{-3}$ fm, equation (\ref{singlemass}) gives that the corresponding axion mass is $m_a \simeq 0.58$ meV. This result has the same magnitude as that presented by \citet{Beck2013}, which gave an axion mass $m_a\sim  0.11$ meV. Considering these, $a\sim 10^{-3}$ fm seems to be a more probable choice of the scattering length for axion or axion-like dark matter particle, though it is not supported by the results of current laboratory experiments. A more detailed discussion about this issue is in the next section.

\section{Discussion and conclusions}\label{sec:discussion}
The BEC axionic scenario of dark matter is becoming more and more prestigious for its fruitful astronomical and cosmological implications. In this paper, we carried an analysis of 17 high-resolution galactic rotation curves in the context of the axion BEC dark matter model. We considered a dilute, axionic BEC gas cloud in each of the sample galaxies. Assuming a repulsive atom-atom interaction, we solved the GP equation for BEC state which is gravitationally trapped. The analytical solution suggested that the size of the BEC cloud has an upper bound which is a function of both the $s$-wave scattering length and the single particle mass $m_a$ of the axions. We then employed the THINGS dataset to put a constraint on the values of the total axion mass $M$ and the maximum possible radius $R$ of the axion BEC cloud. A best-fit value of $R\simeq 39.9$ kpc and $M\simeq 1.3\times 10^{11} M_{\odot}$ have been obtained. We gave the rotation-curve-constrained axion mass density as $\rho_{a}\simeq 0.02~{\rm GeV/cm}^3$. It agrees with the result $\rho_{a}=0.05~{\rm GeV/cm}^3$ presented by \citet{Beck2013} and satisfies the constraint put by \citet{Hoskins2011}.

Besides its mass/number density, the possible axion mass range has been constrained by many experiments of particle physics. Using a superconducting first-stage amplifier (SQUID) to search for dark-matter axions, \citet{Hoskins2011} excluded the possible dark-matter axion mass range of $m_a = 3.3~ \mu {\rm eV}$ to $3.69~\mu {\rm eV}$. This result overlaps with the null range $3.3~\mu {\rm eV}$ to $3.53~\mu {\rm eV}$ presented by \citet{Asztalos2010}. Arik \cite{Arik2013} excluded the possibility of finding dark-matter axions over the mass range $0.64$ eV $\leq m_a \leq$ $1.17$ eV. 
The axion mass we obtained from the fits was given in equation (\ref{singlemass}). It depends on the specific value of $a$. For atomic species $^{23}$Na \cite{Tiesinga1996} and $^{87}$Rb \cite{Boesten1997} used in the laboratory experiments, $a$ is approximately $\sim 10^6$ fm. Equation (\ref{singlemass}) gives that $m_a \simeq 0.58$ eV. Comparing the calculated critical temperature $T_a$ of the axionic BEC dark matter with the ambient temperature for galaxies and galaxy clusters implies that $a\sim 10^{-3}$ fm. This leads to $m_a \simeq 0.58$ meV. It is slightly larger than that presented by \citet{Beck2013}, who gave an axion mass $m_a\sim  0.11$ meV. Both results agree with the constraints on $m_a$ from all the above experiments. These facts implies that $a\sim 10^{-3}$ fm would probably be a more suitable choice for the scattering length $a$, though it is not supported by laboratory experiments nowadays.

Moreover, from equation (\ref{radius}), one can see that there is a parameter degeneracy between $m_a$ and $a$. This means that it is not possible for one to use a least-$\chi^2$ fitting procedure to simultaneously pin down the value of $m_a$ and $a$ but only a combination of them (as the radius $R$). This is the reason that we took $M$ and $R$ (instead of $m_a$ and $a$, which seem to be more ``fundamental'') as free parameters in our numerical analysis. Given a best-fit $R$, one can use equation (27) to determine $m_a$ for different values of $a$.

Another necessary comment on the value of $a$ is the magnitude of the quantity $n|a|^3$. $n$ is the local number density of particles of BEC dark matter given by equation (\ref{number density}). According to Dalfova et al. \cite{Dalfova1999}, the system is said to be dilute or weakly interacting if $n|a|^3 \ll 1$. To estimate $n|a|^3$, one can write it in terms of $M$ and $R$ (using equation (27)), i.e.
\be
n|a|^3=\frac{3M}{4\pi}\left(\frac{G}{\pi^2 \hbar^2 R^7}\right)^{1/3} a^{8/3}.
\ee
Given the best-fit $R\simeq 39.9$ kpc and $M\simeq 1.3\times 10^{11} M_{\odot}$, the condition $n|a|^3 \ll 1$ implies that $a \lesssim 10^{10}$ fm. Assuming $a \simeq 10^{-3}, 1, 10^6, 10^{10}$ fm respectively gives $n|a|^3 \simeq 10^{-38}, 10^{-30}, 10^{-14}, 10^{-3}$. All of these results satisfy the restriction $n|a|^3 \ll 1$.

In fact, $a$ can also take a negative value \cite{Abraham1995}. The scattering length $a$ is a crucial quantity that governing the (non)equilibrium properties of the Bose-Einstein condensate \cite{Boesten1997}. It is generally believed that a positive $a$ guarantees the stability of a homogeneous BEC gas cloud \cite{Fetter1971, Tiesinga1992}. For $a <0$, it is also possible to form a stable BEC in a trap \cite{Ruprecht1995, Kagan1996}. Determination of the value of $a$ is an intricate task and there are lots of nice work in this field \cite{Tiesinga1996,Boesten1997,Abraham1995,Abraham1997,Theis2004}. However, since neither have axions been detected nor has the nature of dark matter been made clear, there is no way to specify the value of $a$ but to leave it for future studies. Investigations in combination with gravitational lensing data and other astronomical observations to estimate $m_a$ and $a$ are currently undertaken.

\section*{Acknowledgments}
We are grateful to Z. Chang and H.-N. Lin from the theory division of Institute of High Energy Physics for useful discussions. We would also like to thank the anonymous editor and referee for their informative comments and constructive suggestions in improving the manuscript.

\newpage

\begin{table*}
\caption{\small{Properties of the sample galaxies from the THINGs dataset. Column (1): galaxy names. Columns (2) and (3): galaxy centers in J2000.0 coordinate from \cite{Walter2008}. Columns (4) and (5): inclinations  and position angles from \cite{Walter2008}. Column (6): systematic velocities from \cite{Blok:2008}. Column (7): distances from \cite{Walter2008}. Column (8): the scale lengths of optical disk from \cite{Blok:2008} and \cite{Mastache:2012ep}. Column (9): mass of HI gas from \cite{Walter2008}. Column (10): apparent B-band magnitudes from \cite{Walter2008}. Column (11): absolute $B$-band magnitudes from \cite{Walter2008}. Column (12): luminosity calculated from column (11) using equation (\ref{eq:M-L-relation}).}}
\begin{ruledtabular}
\begin{tabular}[t]{cccccccccccc}
(1) & (2)& (3) & (4) & (5) & (6) & (7) & (8) & (9) & (10) & (11)  &  (12) \\
 Names   &   RA  &  DEC  &  INCL  &  PA  &  $V_{\rm sys}$ &  $D$   &  $h$ &  $M_{\rm HI}$  & $m_B$  &  $M_B$  &  $L$ \\
 &  [h m s]  &  [$^{\circ}~^{\prime}~^{\prime\prime}$] &  [$^{\circ}$]  &  [$^{\circ}$]  & [km/s] & [Mpc] &  [kpc]  &  $[10^8 M_{\odot}]$ & [mag]  & [mag]  & [$10^{10} L_{\odot}$] \\
\hline
NGC925  & 02 27 16.5 & +33 34 44 & 66 & 287 & 546.3 & 9.2 & 3.30 & 45.8 &9.77 & \---20.04 & 1.614\\
NGC2366 & 07 28 53.4 & +69 12 51 & 64 & 40  & 104.0 & 3.4 & 1.76 & 6.49 &10.51 & \---17.17 & 0.115\\
NGC2403 & 07 36 51.1 & +65 36 03 & 63 & 124 & 132.8 & 3.2 & 1.81 & 25.8 & 8.11 & \---19.43 & 0.920\\
NGC2841 & 09 22 02.6 & +50 58 35 & 74 & 153 & 633.7 & 14.1 & 4.22 & 85.8 & 9.54 & \---21.21 & 4.742\\
NGC2903 & 09 32 10.1 & +21 30 04 & 65 & 204 & 555.6 &  8.9 & 2.40  & 43.5  & 8.82 & \---20.93 & 3.664\\
NGC2976 & 09 47 15.3 & +67 55 00 & 65 & 335 & 1.1   & 3.6 & 0.91 & 1.36 & 9.98 & \---17.78 & 0.201\\
NGC3031 & 09 55 33.1 & +69 03 55 & 59 & 330 & \---39.8 & 3.6 & 1.93 & 36.4 &7.07 & \---20.73 & 3.048\\
NGC3198 & 10 19 55.0 & +45 32 59 & 72 & 215 & 660.7 &  13.8 & 3.06 & 101.7 & 9.95 & \---20.75 & 3.105\\
NGC3521 & 10 05 48.6 & \---00 02 09 & 73 & 340 & 803.5 &  10.7 & 3.09 & 80.2 & 9.21 & \---20.94 & 3.698\\
NGC3621 & 11 18 16.5 & \---32 48 51 & 65 & 345 & 728.5 & 6.6 & 2.61 & 70.7 &9.06 & \---20.05 & 1.629\\
NGC4736 & 12 50 53.0 & +41 07 13 & 41 & 296 & 306.7 & 4.7 & 1.99 & 4.00 &8.54 & \---19.80 & 1.294\\
NGC5055 & 13 15 49.2 & +42 01 45 & 59 & 102 & 496.8 &  10.1 & 3.68 & 91.0 & 8.90 & \---21.12 & 4.365\\
NGC6946 & 20 34 52.2 & +60 09 14 & 33 & 243 & 43.7  & 5.8 & 2.97 & 41.5 &8.24 & \---20.61 & 2.729\\
NGC7331 & 22 27 04.1 & +34 24 57 & 76 & 168 & 818.3 &  14.7 & 2.41 & 91.3 & 9.17 & \---21.67 & 7.244\\
NGC7793 & 23 57 49.7 & \---32 35 28 & 50 & 290 & 226.2 & 3.9 & 1.25 & 8.88 & 9.17 & \---18.79 & 0.511\\
IC2574  & 10 28 27.7 & +68 24 59 & 53 & 56  & 53.1  & 4.0 & 2.56 & 14.8 & 9.91 & \---18.11 & 0.273\\
DDO154  & 12 54 05.8 & +27 09 10 & 66 & 230 & 375.8 & 4.3 & 0.72 & 3.58 & 13.94 & \---14.23 & 0.008\\
\end{tabular}\label{tab:samples}
\end{ruledtabular}
\end{table*}

\begin{table*}
\caption{Results of the three-parameter best fit. The free parameters are the central mass surface density of the stellar disk $\Sigma_0$, the total mass of the axionic BEC dark matter $M$, and the maximum radius of the BEC cloud $R$. $M_{\rm disk}$ is the mass of the disk calculated from equation (\ref{eq:disk-mass}). $M_{{\rm disk}}/L$ is the mass-to-light ratio deduced from equation (\ref{eq:mass-to-light}). $\chi^2/n$ is the reduced chi-square. The numbers after ``$\pm$" are the $1$-$\sigma$ ($68.3\%$ C.L.) errors of the corresponding parameters. Numbers with parenthesis are dismissed for mean values without parenthesis. Mean values with parenthesis are calculated from all the results in the corresponding column.}
\begin{ruledtabular}
\begin{tabular}[t]{ccccccc}
& $\Sigma_0$ & $M$ & $R$ & $M_{\rm disk}$ & $M_{{\rm disk}}/L$ & $\chi^2/n$\\
& $[M_{\odot}~\rm{pc}^{-2}]$ & $[10^{11} M_{\odot}]$ & $[{\rm kpc}]$ & $[10^{10} M_{\odot}]$ & $[M_{\odot}/L_{\odot}]$ & \\
\hline
NGC925 & $21.69\pm 3.86$ & $0.28\pm 0.0043$ & $12.53\pm 0.13$ & $0.15\pm 0.26$ & $0.092\pm 0.016$ & 0.15\\
NGC2366 & $34.39\pm 4.26$ & $0.014\pm 0.002$ & $ 5.60\pm0.16$ & $0.067\pm 0.0083$ & $0.58\pm 0.072$ & 0.26\\
NGC2403 & $516.25\pm 11.33$   & $0.39\pm 0.03$ & $16.46\pm 0.69$ & $1.36\pm 0.03$ & $1.48\pm 0.033$ & 0.29\\
NGC2841 & $2445.8\pm 73.33$ & $4.10\pm 0.24$ & $37.4\pm 1.25$ & $19.37\pm 0.58$ & $4.08\pm 0.12$ & 0.38\\
NGC2903 & $1215.8\pm 46.65$  & $1.2\pm 0.13$  & $25.8\pm 1.77$ & $6.03\pm 0.23$ & $1.65\pm 0.06$ & 0.85\\
NGC2976 & $291.45\pm 17.36$ & $0.24\pm 0.04$ & $12.20\pm 1.05$ & $0.15\pm 0.009$ & $0.75\pm  0.045$ & 0.23\\
NGC3031 & $2642.9\pm (259.03)$ & $0.5\pm 0.12$ & $11.4\pm 1.02$ & $6.19\pm 0.61$ & $2.03\pm 0.20$ & 4.03\\
NGC3198 & $446.34\pm 13.47$ & $1.24\pm 0.09$  & $37.58\pm 1.55$ & $4.22\pm 0.13$ & $1.36\pm 0.04$ & 0.50\\
NGC3521 & $1269.5\pm 27.20$ & $1.7\pm 0.08$  & $31.7\pm 1.01$ & $6.52\pm 0.14$ & $1.76\pm 0.04$ & 0.11\\
NGC3621 & $(526.49)$ & $(280.57)$ & $(219.94)$ & $(2.25)$ & $(1.38)$ & $(0.08)$\\
NGC4736 & $1167.8$ & $5.3$ & $59.5$ & $2.91$ & $2.25$ & $ 8.85$\\
NGC5055 & $1114.6\pm 29.38$ & $1.5\pm 0.69$ & $40.2\pm 10.22$ & $9.48\pm 0.25$ & $2.17\pm 0.057$ & 0.53\\
NGC6946 & $1120.9\pm 30.51$ & $(28.4)\pm (435.76)$ & $98.2\pm (512.50)$ & $6.21\pm 0.17$ & $ 2.28\pm 0.062$ & 0.30\\
NGC7331 & $2236.9\pm (126.27)$ & $3.5\pm 0.18$ & $36.4\pm 1.20$ & $8.14\pm 0.46$ & $1.12\pm 0.06$ & 0.22\\
NGC7793 & $544.30\pm 49.07$ & $0.19\pm 0.032$ & $11.02\pm 1.04$ & $0.53\pm 0.048$ & $1.05\pm 0.094$ & 3.53\\
IC2574 & $26.24\pm 2.45$ & $0.16\pm 0.048$ & $16.61\pm 2.17$ & $0.108\pm 0.01$ & $0.40\pm 0.037$ & 0.11\\
DDO154   & $33.34\pm 4.19$ & $0.026\pm 0.0013$  & $6.34\pm 0.16$ & $0.0061\pm 0.0008$ & $0.76\pm 0.096$ & 0.39\\
\hline
Mean & $945.52\pm 43.66$ & $1.27\pm 0.11$ & $28.69\pm 1.46$ & $4.47\pm 0.18$ & $1.49\pm 0.06$ & 1.30\\
~         & $(920.87\pm 41.10)$ & $(19.37\pm 25.73)$ & $(39.94\pm 31.52)$ & $(4.33\pm 0.17)$ & $(1.48\pm 0.06)$ & $(1.22)$\\
\end{tabular}\label{tab:parameters1}
\end{ruledtabular}
\end{table*}

\nocite{*}


\end{document}